\documentclass[sigconf, screen]{acmart}

\acmConference[ICSE 2024]{46th International Conference on Software Engineering}{April 2024}{Lisbon, Portugal}
\author{Hao Yu}
\affiliation{%
  \institution{School of Software and Microelectronics, Peking University}
  \country{China}
}
\email{yh0315@pku.edu.cn}

\author{Bo Shen}
\affiliation{%
  \institution{Huawei Cloud Computing Technologies Co., Ltd.}
  \country{China}
}
\email{shenbo21@huawei.com}

\author{Dezhi Ran}
\affiliation{%
  \institution{Key Lab of HCST (PKU), MOE; SCS\\Peking University}
  \country{China}
}
\email{dezhiran@pku.edu.cn}

\author{Jiaxin Zhang}
\affiliation{%
  \institution{Huawei Cloud Computing Technologies Co., Ltd.}
  \country{China}
}
\email{zhangjiaxin35@huawei.com}

\author{Qi Zhang}
\affiliation{%
  \institution{Huawei Cloud Computing Technologies Co., Ltd.}
  \country{China}
}
\email{zhangqi98@huawei.com}

\author{Yuchi Ma}
\affiliation{%
  \institution{Huawei Cloud Computing Technologies Co., Ltd.}
  \country{China}
}
\email{mayuchi1@huawei.com}

\author{Guangtai Liang}
\affiliation{%
  \institution{Huawei Cloud Computing Technologies Co., Ltd.}
  \country{China}
}
\email{liangguangtai@huawei.com}

\author{Ying Li}
\authornotemark[1]
\affiliation{%
  \institution{School of Software and Microelectronics, Peking University}
  \country{China}
}
\email{li.ying@pku.edu.cn}

\author{Qianxiang Wang}
\authornotemark[1]
\affiliation{%
  \institution{Huawei Cloud Computing Technologies Co., Ltd.}
  \country{China}
}
\email{wangqianxiang@huawei.com}

\author{Tao Xie}
\authornote{Corresponding author}
\affiliation{%
  \institution{Key Lab of HCST (PKU), MOE; SCS\\Peking University}
  \city{Beijing}
  \country{China}
}
\email{taoxie@pku.edu.cn}

\usepackage{microtype}
\usepackage{graphicx}
\usepackage{subfigure}
\usepackage{booktabs} 
\usepackage{amsmath,amsfonts}
\usepackage{algorithmic}
\usepackage{textcomp}
\usepackage{xcolor}
\usepackage{makecell}
\usepackage{multirow}
\usepackage{diagbox}
\usepackage{balance}
\usepackage[flushleft]{threeparttable}
\usepackage[utf8]{inputenc} 
\usepackage[T1]{fontenc}    
\usepackage{hyperref}       
\usepackage{url}            
\usepackage{booktabs}      
\usepackage{amsfonts}       
\usepackage{nicefrac}      
\usepackage{microtype}
\usepackage{multirow}
\usepackage{listings}
\usepackage{adjustbox}
\usepackage{diagbox}
\usepackage{algorithm}
\usepackage{booktabs}
\usepackage{enumitem}
\usepackage{framed}

\usepackage{pifont}
\newcommand{\codePLM}{PanGu-Coder}

\newcommand{\wenwangbench}{CoderEval}

\newcommand{\newfilejava}{``NewTestFile''}

\setcopyright{acmlicensed}
\copyrightyear{2024} 
\acmYear{2024}
\acmDOI{10.1145/3597503.3623322} 

\acmConference[ICSE 2024]{2024 IEEE/ACM 46th International Conference on Software Engineering}{April 14--20, 2024}{Lisbon, Portugal}
\acmBooktitle{2024 IEEE/ACM 46th International Conference on Software Engineering (ICSE '24),
  April 14--20, 2024, Lisbon, Portugal}
\acmPrice{15.00}
\acmISBN{979-8-4007-0217-4/24/04}

\begin{document}

\title{CoderEval: A Benchmark of Pragmatic Code Generation with Generative Pre-trained Models}

\begin{abstract}
Code generation models based on the pre-training and fine-tuning paradigm have been increasingly attempted by both academia and industry, resulting in well-known industrial models such as Codex, CodeGen, and PanGu-Coder. 
To evaluate the effectiveness of these models, multiple existing benchmarks (e.g., HumanEval and AiXBench) are proposed, including only cases of generating a standalone function, i.e., a function that may invoke or access only built-in functions and standard libraries.
However, non-standalone functions, which typically are not included in the existing benchmarks, constitute more than 70\% of the functions in popular open-source projects, and evaluating models' effectiveness on standalone functions cannot reflect these models' effectiveness on pragmatic code generation scenarios (i.e., code generation for real settings of open source or proprietary code).

To help bridge the preceding gap, in this paper, we propose a benchmark named CoderEval, consisting of 230 Python and 230 Java code generation problems carefully curated from popular real-world open-source projects and a self-contained execution platform to automatically assess the functional correctness of generated code.
CoderEval supports code generation problems from six levels of context dependency, where context refers to code elements such as types, APIs, variables, and consts defined outside the target function  but within the dependent third-party libraries,
current class, file, or project. CoderEval can be used to evaluate the effectiveness of models in generating code beyond only standalone functions.
By evaluating three state-of-the-art code generation models (CodeGen, PanGu-Coder, and ChatGPT) on CoderEval and HumanEval, we find that the effectiveness of these models in generating standalone functions is substantially higher than that in generating non-standalone functions.
Our analysis highlights the current progress and pinpoints future directions to further improve a  model's effectiveness by leveraging contextual information for pragmatic code generation.

\end{abstract}




\maketitle
\section{Introduction}
\label{sec:intro}

Recent years have seen a trend to tackle open-domain code generation tasks with machine learning techniques, especially large generative pre-trained language models~\cite{radford2018improving,brown2020language,gpt-neo,Mikel2019BART} based on Transformer~\cite{vaswani2017attention}, such as Codex~\cite{codex}, AlphaCode~\cite{alpha_code}, InCoder~\cite{fried2022incoder},
CodeGen~\cite{nijkamp2022conversational}, \codePLM{}~\cite{fenia2022pangu}, and ChatGPT~\cite{schulman2022chatgpt}.
Given natural language descriptions specifying the functionalities of the target function, these models can generate both standalone functions (i.e., functions that invoke or access only  built-in functions and standard libraries) and non-standalone functions.

To fairly and comprehensively evaluate the effectiveness of the preceding models, representative benchmarks~\cite{codex, Federico2022multibench,hao2022aixbench,lai2022ds1000,wang2022execution,hendycks-etal-2021-apps,austin2021mbpp,haluptzok2022language,wang2022language,chandel2022training} are required and widely used in the literature.
Released alongside Codex~\cite{codex}, \textbf{HumanEval} is a benchmark for Python to assess the functional correctness of programs generated by code generation models. 
HumanEval consists of 164 hand-written problems, each of which includes a function signature, a docstring, a canonical reference function, and multiple unit tests.
Recently, \textbf{DS-1000}~\cite{lai2022ds1000} is proposed to evaluate the effectiveness of code generation models in generating code that relies on  third-party data science libraries. 
In addition to Python, there are benchmarks for  Java (\textbf{AixBench}~\cite{hao2022aixbench})  and other programming languages (\textbf{MultiPL-E}~\cite{Federico2022multibench}), facilitating the understanding and development of code generation models for these other languages.


Despite the importance and usefulness of these preceding benchmarks, there exists a gap between these benchmarks and pragmatic code generation scenarios (i.e., code generation for real settings of open source or proprietary code). On one hand, \textbf{standalone functions} are heavily focused by these  benchmarks. 
In particular, all the preceding benchmarks except DS-1000 include only standalone functions. Note that DS-1000 includes standalone functions in addition to  non-standalone functions that invoke API functions from only seven specific widely used third-party libraries in data science. 
On the other hand, \textbf{non-standalone functions} commonly exist in pragmatic code generation scenarios. 
After analyzing the 100 most popular projects written in Java and Python, respectively, on GitHub, we find that non-standalone functions account for more than  70\% functions\footnote{Detailed statistics are listed in our open-source repository~\cite{sourcedata}.} of the open-source projects.

To help bridge the preceding gap, 
in this paper, we propose \wenwangbench{}, a \textbf{context-aware} benchmark, which can be used to evaluate code generation models on pragmatic code generation.
According to the source of dependency outside the target function, we categorize code generation problems into six levels (i.e., self\_contained, slib\_runnable, plib\_runnable, class\_runnable, file\_runnable, and project\_runnable), with details described in Section~\ref{sec:bench}.
Given that the existing benchmarks such as HumanEval cover only the first two levels, \wenwangbench{} can provide a more practical and representative evaluation of the effectiveness of code generation models in tasks of pragmatic code generation.
Note that functions belonging to the first two levels correspond to  standalone functions and the others correspond to  non-standalone functions.

To construct \wenwangbench{} as a representative and diverse benchmark for tasks of pragmatic code generation, we select and curate code generation problems from real-world projects in three steps, resulting in 230 problems from 43 Python projects and 230 problems from 10 Java projects.
First, we select functions in open source projects from the most frequent 14 tags and with high star counts on GitHub.
For each function\footnote{In this paper, we use the term of function to refer to both Python function and Java method.}, we extract the original docstring (i.e., the natural language description of the function), the function name and signature, the code implementation, and the corresponding test code (if exists) to form one function-level code generation problem.
Additionally, we analyze the detailed contextual information (where context refers to  code elements such as types, APIs, variables, and consts defined outside the target function, but within the dependent third-party libraries,
current class, file, or project) through program dependence analysis and provide it as all-context (all accessible context) and oracle\_context (the actually used context), so as to conduct a fine-grained evaluation about model effectiveness.
Second, to mitigate data leakage (i.e., the original docstring has a high probability of being used as training data by large language models), we recruit 13 experienced engineers to provide a human-labeled version of description as the second  docstring (i.e., human-labeled docstring) to complement the original docstring. 
Third, to improve the evaluation accuracy, we examine the test coverage of the existing tests provided by each project in \wenwangbench{} and manually write additional tests to achieve high test coverage.



To automatically evaluate code generation models, we  automatically measure the Pass@K metric~\cite{codex} and a newly proposed metric named Acc@K (described in Section~\ref{sec:acc}) to assess generated code.
Since \wenwangbench{} involves functions with contextual dependency and non-primitive types, we build a project-level execution platform to provide a ready runtime environment that can automatically assess the functional correctness of generated code.
We develop this platform based on Docker, where we clone and set up environments for all projects. 
Given multiple solutions generated by a model, the platform can automatically place the generated code in the proper location of the project under consideration.


We conduct a comprehensive evaluation of three state-of-the-art code generation models (CodeGen~\cite{nijkamp2022conversational}, \codePLM{}~\cite{fenia2022pangu}, and ChatGPT~\cite{schulman2022chatgpt}) on \wenwangbench{} and compare the difference with HumanEval. 
In the evaluation, for each model, we analyze the overall and level-wise effectiveness, the ability to correctly utilize contextual information, and the effect of the two different natural language descriptions (i.e., the original and human-labeled docstrings). 
Furthermore, among these models, we compute the overlapping and difference in terms of correctly solved functions.
Through the analysis and comparison, we provide a good understanding of existing models and shed light on future directions and further progress.
Experimental results show that in \wenwangbench{} for Python and \wenwangbench{} for Java, the effectiveness of the three models in generating standalone functions is substantially higher than that in generating non-standalone functions.
Considering that more than 70\% of functions in open-source projects belong to non-standalone functions, improving a  model's ability to consider and use contextual information  is vital for the practical value of this  technology.

In summary, we make the following main contributions:

\begin{itemize}
 
\item 
We point out the limitation of the existing benchmarks through an analysis of the 100 most popular open-source projects written in Java and Python, respectively: the existing benchmarks such as HumanEval typically include only standalone functions, whereas non-standalone functions constitute more than 70\% of functions in the open-source projects.

\item We introduce \wenwangbench{}, a benchmark of pragmatic code generation.
\wenwangbench{} originates from open-source projects from various domains and considers non-primitive types, third-party libraries, and project-specific contextual references. In addition, \wenwangbench{} includes the human-labeled docstring for the target function to complement the original docstring.

\item We evaluate and compare three state-of-the-art code generation  models (CodeGen, PanGu-Coder, and ChatGPT) on \wenwangbench{}. Experimental results indicate three important findings: (1) these models do not work as well on non-standalone functions as on  standalone functions, (2) it is important yet challenging for all these models to generate code with contextual dependency, even for ChatGPT (the most powerful model), and (3) the choice of using the human-labeled docstring vs. the original docstring has an impact on code generation
effectiveness.
\end{itemize}
\wenwangbench{} and all the experimental results are open-sourced~\cite{sourcedata} to continually evolve in the code generation community.

\section{Background}\label{sec:background}

In this section,  we first introduce  existing large language models for code generation.
Then, we introduce the benchmarks  used by existing large language models.





\subsection{Large Language Models for Code Generation}

\codePLM{}~\cite{fenia2022pangu} is a pre-trained language model for the task of text-to-code generation, which is based on the PanGu-$\alpha$ architecture~\cite{pangu_alpha} and a two-stage training strategy.  CodeGen~\cite{nijkamp2022conversational} is a series of conversational text-to-code large language models trained on natural language corpora, corpora of multilingual code (i.e., code written in multiple programming languages), and datasets of Python code. 
Codex~\cite{codex}   is the first work to use large generative pre-trained models to generate complete functions from natural language.
 AlphaCode~\cite{alpha_code}  specializes in programming contests and performs on par with average human developers.
InCoder~\cite{fried2022incoder}  is a unified generation model that can perform program synthesis (through left to right generation) and editing (through padding).
InCoder is trained to generate code files from a large number of code bases with specific friendly licenses, where code regions are randomly masked and moved to the end of each file, allowing code to fill with bidirectional context.
In addition to the above-mentioned models, some other models~\cite{austin2021mbpp, li2023starcoder, luo2023wizardcoder,guo2024deepseek,wei2023magicoder} have been proposed recently.


\subsection{Benchmarks for Code Generation}
Our proposed CoderEval is so far the only benchmark that supports project-level code generation and uses the evaluation metrics with Pass@K, which can validate the functional correctness of the generated code.

Benchmarks that contain project-level functions are important and yet difficult to construct for two main reasons. First, it is necessary to ensure that the selected projects can be compiled and sandboxed to achieve successful execution of different projects. However, it is difficult to successfully build and run many existing open-source projects.
Second, to speed up testing, for the target function, the often large number of its covering test cases from its belonging project needs to be desirably reduced but  statically determining whether a test case covers the target function requires non-trivial  construction of functional dependency diagrams and coverage stub analysis. In addition, when the target function is not covered by the test cases of its belonging  project, the builders of the benchmarks need  to have a deep understanding of the often complex logic of the belonging project before writing high-quality test cases.

Most existing benchmarks (e.g., HumanEval~\cite{codex}, MulitPL-E~\cite{Federico2022multibench}, DS-1000~\cite{lai2022ds1000}, and AiXBench~\cite{hao2022aixbench}) use Pass@K for validating the correctness of a  generated function and they contain only standalone functions. Although DS-1000 can validate the correctness of  generated non-standalone  functions, these non-standalone functions are limited to those that invoke API functions from only seven specific widely used third-party libraries in data science. 

Not being a benchmark, Concode~\cite{iyer2018concode} is a  large dataset that contains over 100,000 functions from approximately 33,000 open-source Java repositories. In Concode, functions from the same repository are split into training, validation, and test sets, so that a deep learning model can be trained based on the training and validation sets, and then be tested on the test set. Each function included in Concode is in the form of a pair: its natural language annotation and code implementation.

Although Concode can support validating the effectiveness of code generation approaches on generating non-standalone functions, there are five  main differences between CoderEval and Concode for highlighting CoderEval's advantages.  (1) CoderEval validates the correctness of the  generated function by executing it, whereas Concode does so based on the text similarity between the generated function and the ground-truth function. (2) CoderEval alleviates the problem of data leakage by providing human-labeled docstrings whereas docstrings in Concode are likely ``seen'' by large language models during the pre-training stage. (3) The functions in CoderEval have been carefully selected by developers, whereas the functions in Concode are automatically collected and have not been carefully selected by developers.  (4) CoderEval supports both Java and Python languages, whereas Concode supports only  Java. (5) CoderEval contains latest code from the past five years, whereas all functions in Concode come from open-source code before 2018. 

\subsection{Statistical Comparison with Different Benchmarks}
\begin{table}
\tiny
\caption{Statistical comparison among CoderEval, HumanEval, and Concode}
\label{tab: statistic comparison for CoderEval HumanEval and Concode}
\begin{adjustbox}{width=1.0\columnwidth}
\begin{tabular}{@{}lrrrrr@{}}
\midrule
\diagbox [width=9em,trim=l] {Benchmark}{Statistical} & Cyclomatic Complexity& Line of Code& \\ 
\midrule
CoderEval-Python   & 4.71 & 32.0\\
CoderEval-Java  & 3.10  & 10.2\\
HumanEval  & 3.62  & 7.8\\
Concode  & 1.43  & 4.8\\
\midrule
\end{tabular}
\end{adjustbox}
\end{table}

Table~\ref{tab: statistic comparison for CoderEval HumanEval and Concode} shows the average value of cyclomatic complexity and average number of lines of code in CoderEval, HumanEval, and Concode. We find that the cyclomatic complexity of functions in CoderEval is similar to HumanEval, and the average number of lines of code in CoderEval is more than that in HumanEval. 
\section{\wenwangbench{} Benchmark}
\label{sec:approach}

\begin{figure*}
	\centering
	\includegraphics[width=1.0\linewidth]{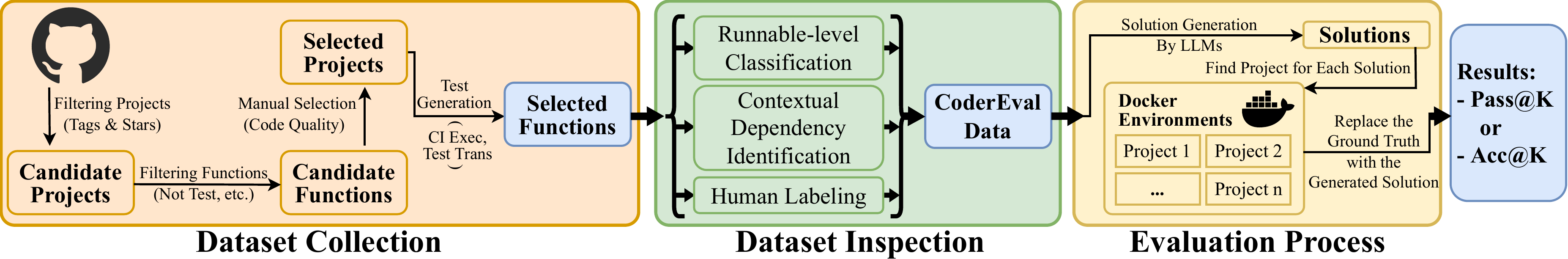}
	\caption{Overview of \wenwangbench{} construction process}
	\label{fig.overview}
 \vspace{-0.2cm}
\end{figure*}

In this section, we introduce \wenwangbench{}, whose construction process is shown in 
Fig~\ref{fig.overview}. The construction process of \wenwangbench{} includes three phases: dataset collection, dataset inspection, and evaluation process.

\subsection{Dataset Collection}


\subsubsection{Problem  Selection.} To make \wenwangbench{} pragmatic and diverse, we select functions from various open-source projects by four steps. 
(1) We select candidate projects by crawling the tags of all projects on GitHub and selecting projects with the most frequent 14 tags and with high stars.
For each tag, we select the projects with the top five highest number of stars.
The 14 types of tags are ``gson'', ``music'', ``logging'', ``chat'', ``websocket'', ``mvc'', ``leetcode'', ``microservices'', ``jdbc'', ``json'', ``crud'', ``datastructures'', ``log4j'', and ``serialization''.
(2) We extract all functions in the selected projects, and keep only the ones that are not a test, interface, or deprecated function, with a function-level comment in English, and can run successfully in the verification platform (Section~\ref{sec.platform}) and pass the original test cases.
(3) We select high-quality functions from the candidate functions through manual screening, whose main criterion is whether a function often appears in real development scenarios.
(4) We attain projects according to the number of the selected functions contained in each project. This process can help us compile fewer projects with the same number of total selected functions.
In the end, we attain 230 problems from 10 projects in Java.
To maintain consistency with the number of problems in CoderEval for Java, we also attain 230 problems for CoderEval for Python from 43 projects.

The reason why we select high-quality functions manually (the third step described above) is that we want CoderEval to assess a model's ability to generate code that is helpful to developers. 
We engage experienced developers to select functions that may be used in real scenarios, following five rules.  
(1)	Functions that contain fewer than ten contextual tokens. Excessive contextual dependencies make it difficult for a model to generate correct solutions. The functions that are too difficult for all LLMs to generate correct implementations are not suitable for the benchmark.
(2)	Functions that are frequently used in real development scenarios judged by any of the 13 experienced engineers recruited by us. The rationale behind this rule is that different developers have various preferences over  frequently used functions.
(3)	Functions containing docstrings that can reflect the implementations of the functions.
(4)	Functions that have more than three lines of code implementation.
(5)	Functions that are not test or deprecated functions.




\subsubsection{Test Construction.} For each project included \wenwangbench{}, we first run the unit test cases contained in the original project and measure the branch coverage of each function (from the project) included in \wenwangbench{}. When a function's branch coverage has not achieved 100\%, the first author of this paper has manually written additional test cases for this function to aim to achieve 100\% branch coverage (if ever possible).

Based on the unit test cases in the original projects (in addition to those manually written by the first author), we construct test cases in \wenwangbench{} via two steps. 
(1) To obtain the corresponding test cases of a function in \wenwangbench{}, we construct a static function call graph from source code and then use the graph to select all the test cases that can reach the function. In doing so, when we assess the correctness of a function, we avoid the high cost of executing all the test cases in the function's belonging project.
(2) To unify and simplify the interface of invoking the test cases in \wenwangbench{}, we automatically convert the corresponding test cases (typically depending on test frameworks such as JUnit and TestNG) of a function in \wenwangbench{} into non-test functions with the ``main'' function in a new Python/Java file. 
We denote these new Python/Java files as \newfilejava{}.

\begin{table*}[t]
\centering
\caption{Dependencies belonging to each contextual dependency type}
    \renewcommand{\arraystretch}{1.3}
    \begin{tabular}{llll}
    \toprule
    \multicolumn{1}{c}{Dependency Type\tnote{1}} & \multicolumn{1}{c}{Dependencies} & \multicolumn{1}{c}{Examples in Python} & \multicolumn{1}{c}{Examples in Java}\\ 
    \midrule
    self-contained & built-in types/functions, no need to import & min(),  print() & System.xxx\\
    slib-runnable & standard libraries/modules, no need to install & os, subprocess, sys & Arrays.sort()\\
    plib-runnable & publicly available libraries on pypi/maven & unittest2, requests & com.google.code.gson\\
    class-runnable & code outside the function but within class & self.xxx,  X.f() & this.f()\\
    file-runnable & code outside the class but within the file & func(), URL, name & func()\\
    project-runnable & code in the other source files & superclass, utils & superclass, utils\\
     \bottomrule
    \end{tabular}
        
        Each runnable level  must depend on the dependencies defined at this level,  must not depend on the
        
        dependencies defined at its subsequent levels, and may or may not depend on dependencies defined at  its previous levels.

    \label{tab: run_dep_cla}
\vspace{-0.2cm}
\end{table*}

\subsection{Dataset Inspection}

\subsubsection{Human-labeled Docstring.} When large language models (LLMs) are used for code generation, choosing what prompts to use can have a great impact on the quality of the generated code. To study the effect of different prompts, we recruit 13 experienced engineers with at least 3 years experience of Python/Java, and let them provide a human-labeled version for the description (named as docstring) of each problem in \wenwangbench{}.

There are three main reasons for including human-labeled docstrings in CoderEval. 
(1) Including human-labeled docstrings helps mitigate the memory effect (i.e., original docstrings have a high probability of being seen by LLMs in the pre-training stage) of LLMs and explore the impact of prompts with human labeling on the quality of code generated by LLMs. (2) Including human-labeled docstrings helps study how LLMs  perform on different docstrings. 
(3) Including human-labeled docstrings helps provide high-quality docstrings for the functions in \wenwangbench{}. During our initial study, we find that the original docstrings (from the selected open-source projects) for the problems in \wenwangbench{} are highly diverse, well beyond describing functionalities, which are typically used as prompts for LLMs.
An original docstring can play multiple different roles, such as explaining the internal logic, introducing the external usage and behavior, declaring the effect and caution.

\subsubsection{Contextual Dependency Identification.}\label{context identify}
One of major differences between HumanEval~\cite{codex} and \wenwangbench{} is that \wenwangbench{} considers the target function's contextual dependency, which refers to code elements defined outside of the target function but required by it in order to run. Table~\ref{tab: run_dep_cla} shows the dependencies belonging to each dependency type along with examples.



We identify the contextual dependencies of a function through program analysis of its belonging project with three steps\footnote{
Note that we also use the same  three steps of identifying contextual dependencies to calculate the proportions of standalone functions and non-standalone functions, respectively, in the 100 most popular projects written in Java and Python on GitHub, respectively (as reported in Section~\ref{sec:intro}).}.
(1) Before the analysis, we first build a knowledge base that helps identify different references of Python/Java builtins and different imports of standard or public libraries from the project-specific code.
To do so, we cache all built-in types/functions/variables/constants and standard library names for each Python version from 3.0.0 to 3.10.0 and Java version from 1.8 to 17, as well as all publicly available libraries on $pypi.org$ and the $Maven$ central repository.
(2) With the knowledge base, given the function under analysis, we parse the source file that contains the function to get a list of type/function/variable/constant definitions.
 (3) We employ static program analysis to identify all the dependent elements namely the external references and invocations whose definitions are outside the function under analysis, and classify them into three categories: type\_reference, variable\_reference, and API invocation. More specifically, type\_reference refers to a user-defined class or a standard type (e.g., ``List'', ``subprocess'', and ``os'').
Variable\_reference refers to a user-defined variable or object.
API invocation refers to a user-defined function or  function in a standard or third-party library.

We use a function's \emph{oracle\_context} information to denote its contextual dependencies identified by the preceding three steps. 
Oracle\_context information not only can be used to evaluate the accuracy of the contextual dependencies in the code generated by an LLM (as done in our experiments) but also can be used as part of the input to the LLM in the inference stage, to explore prompting the LLM additionally with the oracle\_context  information to improve code-generation  effectiveness.

However, it is usually difficult for the users of an LLM to know oracle\_context  information in advance, so \wenwangbench{} also includes  a function's  \emph{all\_context} information  (i.e., all types, variables, and APIs defined/imported in the function's belonging file), which can be additionally used to prompt the LLM
 to generate code incorporating part or all of the all\_context information. 
Note that we do not use the all\_context information in the inference stage during our experiments; there we use only the original docstring, human-labeled docstring, function name, and function signature of the target function.
We hope that the all\_context information can be valuable for the research community, and we plan to study the impact of all\_context information (as part of the used prompt) on code-generation effectiveness in our future work.

\subsubsection{Runnable Level Classification.} As shown in Table~\ref{tab: run_dep_cla}, we classify the target function into six runnable levels, each of which refers to the scope that the function  can run successfully. These levels include  \textit{self-contained}, \textit{slib-runnable}, \textit{plib-runnable}, \textit{class-runnable}, \textit{file-runnable}, and \textit{project-runnable}.
Each runnable level must depend on the dependencies defined at this level,  must not depend on the dependencies defined at its subsequent levels, and may or may not depend on dependencies defined at its previous levels. 
For example, a function at the plib-runnable level depends on public libraries, so it can run successfully outside its belonging class, source file,  and project, as long as the required Python version is used and the libraries are installed and imported.

\begin{figure}
	\centering
	\includegraphics[width=1.0\linewidth]{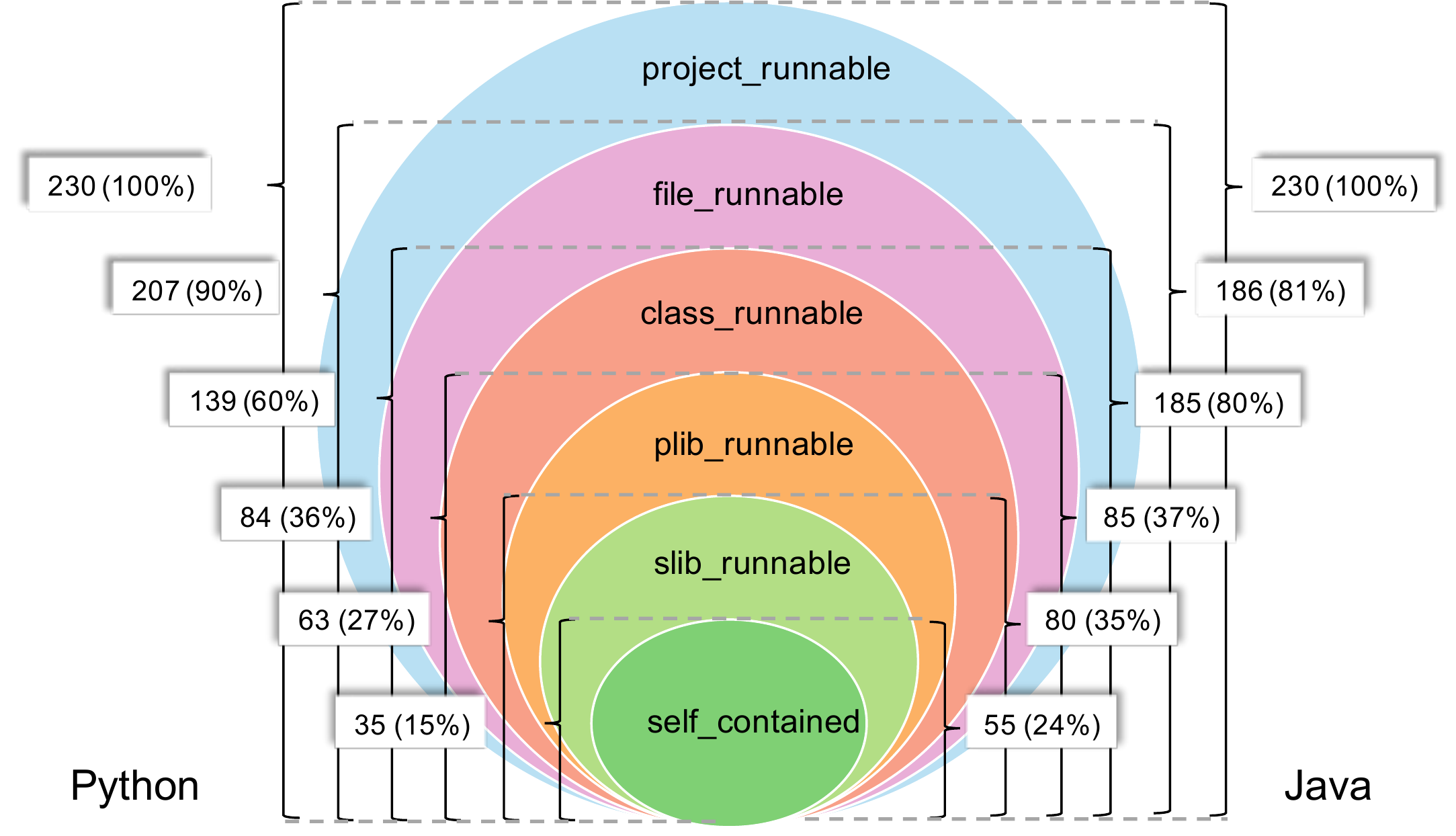}
	\caption{Distribution of runnable levels in \wenwangbench{}}
	\label{fig.run_dep_res}
\vspace{-0.2cm}
\end{figure}

The left and right parts of Fig~\ref{fig.run_dep_res} show the distribution of runnable levels on CoderEval for Python and Java, respectively.
Most (84\%) of the problems collected in \wenwangbench{} for Python are file-runnable,  indicating that the oracle\_context information in their respective belonging  file is necessary and critical for an LLM to generate the target function correctly.
The class-runnable functions account for about half (54\%) of all the functions; this result is not surprising considering the prevalence of object-oriented programming in open-source  non-trivial projects.
On the contrary, the slib-runnable functions account for only 17\% of the functions in CoderEval for Python but 100\% in the case of  HumanEval. This result shows  limitation of HumanEval when used for  model optimization and evaluation, especially in  pragmatic code generation  in real settings.
Similar to CoderEval for Python, the slib-runnable functions account  for only 35\% of the functions in CoderEval for Java.

\subsection{Evaluation Process}\label{sec.platform}

To provide a ready runtime environment to automatically  execute and evaluate a function generated by a model, we construct an evaluation platform based  on a Linux Docker image, which can provide a virtual sandbox to enable easy and safe  distribution.


\subsubsection{Evaluation for Python.} To evaluate a generated Python function for the target function, we need to clone and set up environments for all the Python projects in \wenwangbench{}.
To avoid conflicts of Python/library versions, under each repository's root directory, we first use \emph{pyenv} to set the local Python version to the latest version specified in the CI configuration or document, and then use \emph{venv} to create an individual virtual environment for the target function's belonging project.
After that, we use \emph{pip} to install all the dependencies and trigger execution of the test cases in the original belonging project to ensure successful setup of the runtime environment.

After the runtime environment is successfully set up, our platform  automatically replaces the target function (in its belonging project) with the generated function,  invokes the generated function with the test inputs from the corresponding test cases of the target function, and compares the actual test outputs with the expected outputs from the corresponding test cases to determine whether all the corresponding test cases pass or not. 

\subsubsection{Evaluation for Java.} Similar to the evaluation platform for Python, the evaluation platform for Java  also first clones all the Java projects in \wenwangbench{}.
We ensure that all Java projects in CoderEval can be executed with the Java 17 version.
Different from Python code, Java code needs to be compiled before execution, so the platform  automatically compiles the test file (i.e., \newfilejava{}) (including the corresponding test cases of the target function) in advance, and then replaces the target function with the generated function in the belonging project.

The platform then runs the ``javac'' command to incrementally compile the changed files (whose changes are caused by the function replacement), and runs the ``java'' command to execute the bytecode of \newfilejava{}. The return value of the ``java'' command indicates whether all the corresponding test cases of the target function pass or not. 
Note that our platform uses  the parameter ``-cp'' of both the ``javac'' and ``java'' commands to import the class file of the given test file and its dependent files/libraries. In addition, if the compilation fails, our platform (without executing test cases) determines that not all the corresponding test cases of the target function pass.




\section{Experimental   Setup}\label{sec:evaluation}
\begin{table*}[t]

    \centering
    \caption{Overall effectiveness of three models on two benchmarks}
    \label{tab: overall performance for wenwangbench and he new}
    \begin{threeparttable}
    \begin{tabular}{@{} rr|rrr|rrrr @{}} 
        \toprule
        \multicolumn{1}{c}{\multirow{3.8}{*}{Benchmark}} & \multicolumn{1}{c}{\multirow{3.8}{*}{Model}}
        & \multicolumn{3}{c}{Python} & \multicolumn{3}{c}{Java} &\\
        \cmidrule(lr){3-5}  \cmidrule(lr){6-8}
        &  & Pass@1 & Pass@5 & Pass@10  & Pass@1 & Pass@5 & Pass@10\\
        \midrule
        \multirow{3}{*}{\wenwangbench{}} & CodeGen\tnote{1} &     9.48\% & 19.58\% & 23.48\%    & 13.91\% & 27.34\% & 33.48\%\\
        & \codePLM{}\tnote{2}   &   11.83\% & 20.93\% & 27.39\%        & 25.43\% & 37.39\% & 43.04\% \\
        & ChatGPT\tnote{3}   & 21.04\% & 27.31\% & 30.00\%   & 35.39\% & 42.77\% & 46.09\%\\
        \midrule
        \multirow{3}{*}{HumanEval} & CodeGen\tnote{1}      & 10.20\% & 19.79\% & 22.80\%     & 5.78\% & 9.00\%  & 9.45\% \\
        & \codePLM{}\tnote{2}     & 13.42\% &21.48\% & 22.73\%  & 8.21\% & 15.88\% & 18.63\%  \\
        & ChatGPT\tnote{3}   & 39.21\%& 64.09\% & 72.96\%  & 38.21\% & 59.35\% & 67.23\% \\
        \bottomrule
    
    \end{tabular}
        \begin{tablenotes}
        \footnotesize
        \item[1] We use the 350M CodeGen-Mono model with the default settings for Python. Since CodeGen does not have a monolingual version of Java, on \wenwangbench{} for Java, we use the CodeGen-Multi model instead.
        \item[2] \codePLM{} has two different models for Python and Java, and we use the 300M models with the default settings for Python and Java. 
        \item[3] We use the ``gpt-3.5-turbo'' for ChatGPT in our experiments.
        
      \end{tablenotes}
\end{threeparttable}
      \vspace*{-2.5ex}
\end{table*} 

In this section, we describe the setup of our experiment with three models (CodeGen~\cite{nijkamp2022conversational}, \codePLM{}~\cite{fenia2022pangu}, and ChatGPT~\cite{schulman2022chatgpt}) on two benchmarks (\wenwangbench{} and HumanEval~\cite{codex}) in terms of research questions, model settings, and  evaluation metric.
To alleviate the randomness of a single run of experiment, we run each experiment 10 times and find that the standard deviation of the used  evaluation metric (namely Pass@K) for all models is about 1\%. To better show the experimental results (e.g., complementarity and overlapping of different models in generating correct functions), in this section, the discussed results focus on the median group of experiments from 10 experiments according to the Pass@10 metric value.
Our experimental results and related artifacts are publicly available~\cite{sourcedata}.
\subsection{Research Questions}

Our experiment intends to answer the following research questions:
\begin{itemize}
    \item \textbf{RQ1:} How do the models of CodeGen, \codePLM{}, and ChatGPT perform, especially in generating 
standalone vs. non-standalone functions?

    \item \textbf{RQ2:} How do these models perform in correctly incorporating the oracle\_context information in the generated code?
    
    \item \textbf{RQ3:} How do different prompts impact the effectiveness of these models?

\end{itemize}


\subsection{Model Selection and Settings}
\subsubsection{Model selection}
We focus on models that support code generation for both Python and Java. CodeGen, PanGu-Coder, and ChatGPT all support both Java and Python. Specifically, we choose CodeGen because CodeGen has specialized models for the Python language (CodeGen-mono) and models that support multiple languages (CodeGen-multi). We choose PanGu-Coder because PanGu-Coder has two models specifically designed for Python and Java, respectively. We choose ChatGPT because ChatGPT is the most effective and parameter-intensive multilingual model. Selecting these models helps compare the effectiveness of single-language generative models and multi-language generative models.
The reason why we do not select CodeGeex~\cite{zheng2023codegeex} is that its model is of 13B size, being too large for us to complete the experiments due to our limited computing resources at the time of the paper writing. We plan to evaluate the effectiveness of CodeGeex and other recent open-source models (e.g., StarCoder~\cite{li2023starcoder} and WizardCoder~\cite{luo2023wizardcoder}) on CoderEval in future work.
\subsubsection{Model settings}
For \codePLM{}, we use the 300M \codePLM{} model with the default settings.
For CodeGen, we use the 350M CodeGen-Mono model with the default settings for Python.
Since CodeGen does not have a monolingual version of Java, on \wenwangbench{} for Java, we use the CodeGen-Multi model instead.
For ChatGPT, we use the ``gpt-3.5-turbo'' in our experiments.
ChatGPT’s scale of model parameters is much larger than the other two models.

In the inference stage, for all models, we set the max window length to 1024.
We use nucleus sampling~\cite{Holtzman2020sample}: the number of samples is 10 (i.e., generating 10 codes per function), and the temperature is 0.8.

\subsection{Evaluation Metric}\label{sec:acc}
Similar to HumanEval, we adopt the Pass@K metric for an LLM. In particular, given the 230 problems in \wenwangbench{}, Pass@K for an LLM measures the percentage of the problems (among the 230 problems) for which there is at least one correctly (judged based on running the corresponding test cases) generated solution among the top K samples (i.e., solutions) generated by the LLM. 
During the experiments, we set the total number (denoted as  \emph{n}) 
 of samples generated by an LLM to 10, and then calculate Pass@K for the LLM with K's value of 1, 5, and 10, respectively. To avoid the issue of high sampling variance, we use the unbiased estimator of Pass@K implemented by ChatGPT in HumanEval\footnote{https://github.com/openai/human-eval/}.
Note that although \emph{n} is  set to 200 in previous work on \codePLM{}~\cite{fenia2022pangu} and CodeGen~\cite{nijkamp2022conversational}, we set \emph{n} to 10 as what the Copilot plugin does because we consider that sampling 10 times is more feasible and reasonable at an acceptable cost and response time in practical settings.

Besides Pass@K, we also propose the Acc@K metric for an LLM based on oracle\_context tokens (i.e., dependent elements included in the  oracle\_context information) to evaluate 
the LLM's capability of generating each individual oracle\_context token among the top K samples generated by the LLM. 
In particular, Acc@K for an LLM measures the percentage of the target functions (among the 230 target functions as ground truth solutions for the 230 problems) whose each oracle\_context token  
is included by at least one of K samples generated by the LLM.  
Note that other metrics can be formulated to evaluate an LLM's capability of generating a function including all the oracle\_context tokens in the target function (with and without consideration of the tokens' sequential order, respectively). These metrics impose more challenging requirements for an LLM to satisfy, and we plan to explore these and other metrics in future work.

\section{Experimental Results}\label{sec:experiments}
In this section, we detail the experimental results for CodeGen, PanGu-Coder, and ChatGPT.

\subsection{RQ1: How Do CodeGen, \codePLM{}, and ChatGPT Perform, Especially in Generating 
Standalone vs. Non-Standalone Functions?}

\subsubsection{Overall effectiveness.}\label{sec:overall}
Table~\ref{tab: overall performance for wenwangbench and he new} reports the overall effectiveness of CodeGen, \codePLM{}, and ChatGPT on \wenwangbench{} and HumanEval. 
On HumanEval, the reported effectiveness of the three models on HumanEval for Python is the effectiveness reported in these models' respective corresponding papers.
These papers do not report the effectiveness (being evaluated in our experiments) of their models on HumanEval for Java. 
Since the problems in CoderEval for Python and those in CoderEval for Java do not have overlapping, we cannot directly compare the effectiveness of the three models on \wenwangbench{} for Python and that on \wenwangbench{} for Java.
On both HumanEval and \wenwangbench{} (for both Python and Java), ChatGPT consistently outperforms the other two models.
The reason is that ChatGPT's parameter scale is much larger than the other two models. 
CodeGen and PanGu-Coder perform worse on HumanEval than on CoderEval because although the functions in HumanEval are standalone functions (i.e., a function that may invoke or access only built-in functions and standard libraries), their complexity is higher than that of the standalone functions in CoderEval. 
The functions in HumanEval tend to be more algorithmic, while the standalone functions in CoderEval tend to be more simple and practical (e.g., converting  integer to string or floating point number). 
The effectiveness of ChatGPT in HumanEval is higher than that in CoderEval. ChatGPT's powerful algorithmic learning abilities make ChatGPT perform well on HumanEval. However, when generating code that relies on oracle\_context tokens, the prompts given to ChatGPT do not include all\_context tokens, resulting in poor effectiveness on CoderEval.

\begin{figure}
\centering
\subfigure[\wenwangbench{} for Python]{
\label{fig.complenment_overall_python}
\begin{minipage}[t]{0.5\linewidth}
\centering
\includegraphics[width=\linewidth]{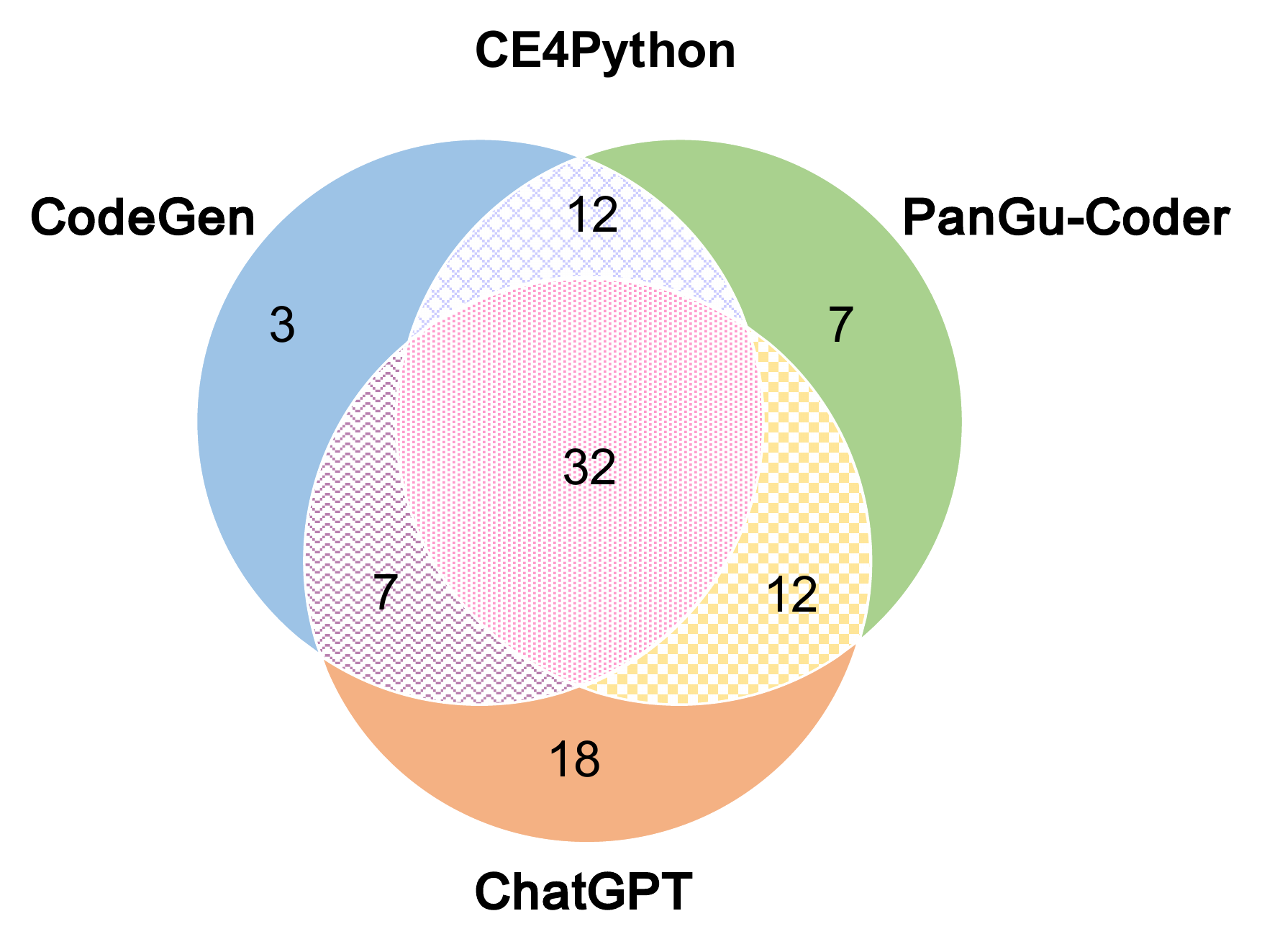}
\end{minipage}%
}%
\subfigure[\wenwangbench{} for Java]{
\label{fig.complenment_overall_java}
\begin{minipage}[t]{0.5\linewidth}
\centering
\includegraphics[width=\linewidth]{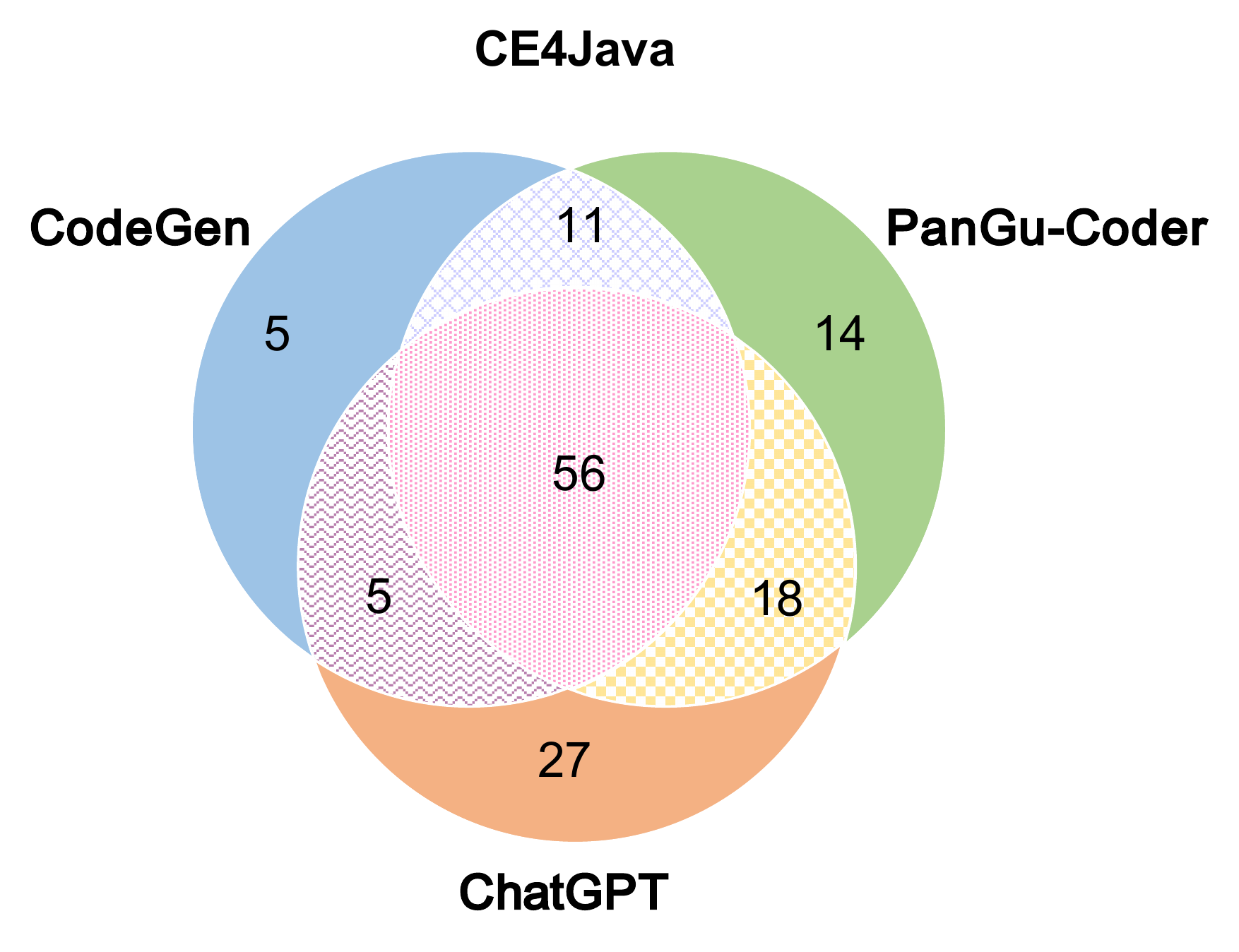}
\end{minipage}%
}%
\centering
\caption{Number of  problems solved by CodeGen, \codePLM{}, and ChatGPT, respectively, on \wenwangbench{}}
\label{fig.complenment_overall}
\end{figure}

\begin{table*}[t]
    \centering
    \caption{Effectiveness comparison between the generation of standalone and non-standalone functions on CoderEval}
    \label{tab: mult performance on wenwangbench}
    \begin{tabular}{@{} r|r|rrr|rrrr @{}} 
        \toprule
        \multicolumn{1}{l|}{\multirow{3}{*}{Runnable-Level}} & \multicolumn{1}{c|}{\multirow{3}{*}{Model}}
        & \multicolumn{3}{c|}{Python} & \multicolumn{3}{c}{Java} &\\
        \cline{3-5}  
        \cline{6-8}
\cline{3-8}
         & & Pass@1 & Pass@5 & Pass@10 &  Pass@1 & Pass@5 & Pass@10\\
         \midrule
        \multicolumn{1}{l|}{\multirow{3}{*}{\textbf{standalone:}}} &  CodeGen  & 18.10\% & 33.38\% & 38.10\% & 26.25\% & 46.40\% &  52.50\%\\
        & \codePLM{} &19.52\% &31.30\% & 38.10\% & 43.88\% & 57.23\% & 62.50\%\\
        & ChatGPT &35.87\% & 43.56\% & 47.62\% & 64.88\% & 68.65\% &  70.00\%\\
\cline{2-8}
        \multicolumn{1}{l|}{\multirow{3}{*}{- self-contained}} &  CodeGen   &  22.57\% & 36.75\% & 40.00\% & 27.27\% & 49.37\% &  56.36\%\\
        & \codePLM{} & 23.71\% &34.93\% & 40.00\% & 48.36\%&  57.94\%	&  61.82\%\\
        & ChatGPT &52.29\% & 60.08\% & 82.86\% & 61.82\% & 66.21\% &  67.27\%\\
        \cline{2-8}
        \multicolumn{1}{l|}{\multirow{3}{*}{- slib-runnable}} &  CodeGen  & 12.50\% & 29.17\% & 35.71\% & 24.00\% & 39.87\% &  44.00\%\\
        & \codePLM{} &  14.29\% & 26.76\% & 35.71\% & 34.00\%	&  55.68\%	&  64.00\%\\
        & ChatGPT & 15.36\% & 22.92\% & 28.57\% & 71.60\% & 74.00\% &  76.00\%\\
\midrule
        \multicolumn{1}{l|}{\multirow{3}{*}{\textbf{non-standalone:}}} &  CodeGen  & 6.23\% & 14.38\% & 17.96\% & 7.33\% & 17.17\% &  23.33\%\\
        & \codePLM{} &  8.92\% & 17.02\% & 23.35\% & 15.60\% & 26.81\% & 32.67\%\\
        & ChatGPT &15.45\% & 21.18\% & 23.35\% & 19.67\% & 28.97\% &  33.33\%\\
\cline{2-8}
        \multicolumn{1}{l|}{\multirow{3}{*}{- plib-runnable}} & CodeGen & 4.76\% & 16.16\% & 23.81\% & 0\% & 0\% & 0\%\\
        & \codePLM{} & 13.33\% & 22.75\% & 28.57\% & 0\% & 0\% & 0\%\\
        & ChatGPT & 21.43\% & 28.06\% & 28.57\% & 0\%	&  0\%	&  0\%\\
        \cline{2-8}
        \multicolumn{1}{l|}{\multirow{3}{*}{- class-runnable}} & CodeGen & 5.82\% & 9.05\% & 10.91\% & 8.30\% & 19.21\% &  26.00\%\\
        & \codePLM{} & 7.82\% & 15.04\% & 21.82\% & 19.90\%	&  32.59\%	&  40.00\%\\
        & ChatGPT & 8.73\% & 12.57\% & 14.55\% &22.40\% & 31.49\% &  36.00\%\\
        \cline{2-8}
        \multicolumn{1}{l|}{\multirow{3}{*}{- file-runnable}} & CodeGen &  7.79\% & 20.19\% & 25.00\% & 0\% & 0\% & 0\%\\
                & \codePLM{} & 9.41\% & 19.04\% & 26.47\% & 0\% & 0\% & 0\%\\
        & ChatGPT &  21.03\% &  29.09\% &  32.35\% & 0\% & 0\% & 0\%\\
        \cline{2-8}
        \multicolumn{1}{l|}{\multirow{3}{*}{- project-runnable}} & CodeGen & 3.91\% & 8.33\% & 8.70\% & 6.14\% & 14.89\% &  20.45\%\\
        & \codePLM{} & 6.09\% & 10.51\% & 13.04\%& 7.95\%	&  17.33\%	&  20.45\%\\
        & ChatGPT & 9.57\% & 12.08\% &  13.04\% & 16.14\% & 27.20\% &  31.82\%\\
        \bottomrule
    \end{tabular}
    \vspace{-0.2cm}
\end{table*}

\subsubsection{Complementarity and overlapping of three models.} Fig~\ref{fig.complenment_overall} shows the number of solved problems by the three models on \wenwangbench{}, with the difference and overlapping.
We use the median group of experiments selected from 10 experiments with the Pass@10 value to count the passing problems of each model.
We consider a problem to be solved by a model if the model can generate \emph{at least} one  solution (out of the 10 generated solutions) that passes all the test cases.
Among the 230 problems in \wenwangbench{} for Python, the total number of functions solved by at least one of the three models is 91 (complementarity), while 32 problems (overlapping) can be solved by any of the three models.
On \wenwangbench{} for Java, the total number of problems solved by at least one of the three models is 136 (complementarity), while 56 problems (overlapping) can be solved by any of the three models.
The high complementarity indicates that different models have their unique capabilities, and the high overlapping indicates that the ability of the three models to generate the correct problem is consistent in some problems.
We manually analyze the overlapping problems and find that more than half of them belong to standalone functions. More specifically, the total  number of overlapping problems in CoderEval is 88, of which 52 belong to standalone functions. Note that the standalone functions account for only about 30\% in CoderEval.
Therefore, the three models are all good at generating standalone functions (while ChatGPT performs particularly well); extending the ability to non-standalone functions and exploring how to combine the code generation capabilities of different models are worthwhile research directions.

\subsubsection{Effectiveness comparison between the generation of standalone and non-standalone functions.}

Since all functions in HumanEval are standalone functions, we further compare the effectiveness of three models in generating the standalone functions in \wenwangbench{} with the effectiveness of the non-standalone functions in \wenwangbench{}.
The \textit{standalone} and \textit{non-standalone} rows in Table~\ref{tab: mult performance on wenwangbench} show that, in both \wenwangbench{} for Python and \wenwangbench{} for Java, the effectiveness of the three models in generating the standalone functions is substantially higher than that of the non-standalone functions.
Considering that more than 70\% of functions in open-source projects belong to non-standalone functions, improving a model's ability to correctly generate oracle\_context information  is vital for pragmatic code generation.

We evaluate both standalone and non-standalone functions at detailed runnable levels.
Standalone functions include two runnable levels, \textit{self-contained} and \textit{slib-runnable}, and non-standalone functions include four runnable levels, \textit{plib-runnable}, \textit{class-runnable}, \textit{file-runnable}, and \textit{project-runnable}.
From Table~\ref{tab: mult performance on wenwangbench}, we find that except for the Pass@5 and Pass@10 of \textit{class-runnable} on CoderEval for Java, ChatGPT outperforms the other two models on all runnable levels. 
As discussed in Section~\ref{sec:overall}, the main reason is that ChatGPT's parameter scale is much larger than the other two models.

\begin{framed}
In summary, on both CoderEval for Python and CoderEval for Java, the effectiveness of the three models in generating standalone functions is substantially higher than that in generating non-standalone functions. Different models have their unique capabilities in code generation, and how to combine the code generation capabilities of different models is a worthwhile research direction.
\end{framed}

\subsection{RQ2: How Do These Models Perform in Correctly Incorporating the Oracle\_context Information in the Generated Code?}
To further study the ability of LLMs to generate oracle\_context tokens, we analyze the oracle\_context tokens in generated solutions and focus on three categories of tokens: TypeReference, APIInvocation, and VarReference (note that VarReference does not include variables defined in the target function).

\begin{figure*}
	\centering
	\includegraphics[width=1.0\linewidth]{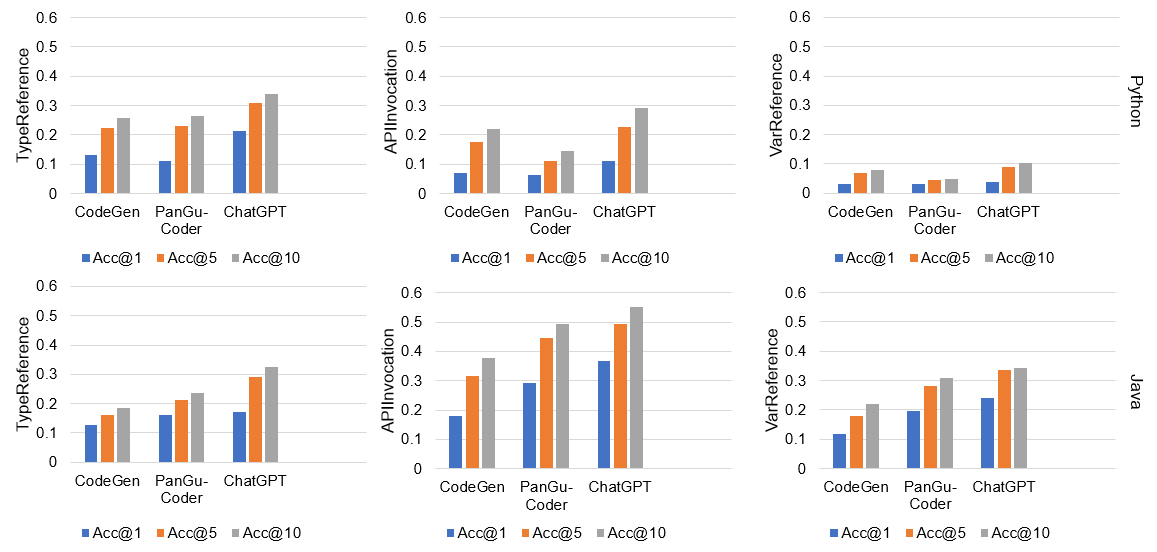}
	\caption{The three models' accuracy of generating the oracle\_context tokens}
	\label{fig.contextual}
 \vspace{-0.2cm}
\end{figure*}
\begin{table}
\tiny
\caption{The similarity between the original docstring and human-labeled docstring}
\label{tab: similarity between human and build-in doc}
\begin{adjustbox}{width=1.0\columnwidth}
\begin{tabular}{@{}rrrrrr@{}}
\midrule
\diagbox [width=7em,trim=l] {Language}{Metrics} & BLEU-4& Jaccard Similarity& \\ 
\midrule
Python   & 54.7 & 55.2\\
Java  & 18.1  & 26.9\\
\midrule
\end{tabular}
\end{adjustbox}
\end{table}

\begin{table*}[t]
    \centering
    \caption{Effectiveness with two prompt versions of \wenwangbench}
    \label{tab: overall performance for wenwangbench with human label}
    \begin{tabular}{@{} rr|rrr|rrrr @{}} 
        \toprule
        \multicolumn{1}{c}{\multirow{3.7}{*}{Model}} & \multicolumn{1}{c}{\multirow{3.7}{*}{Prompt}} 
        & \multicolumn{3}{c}{Python} & \multicolumn{3}{c}{Java} &\\
        \cmidrule(lr){3-5}  \cmidrule(lr){6-8}
        
        & &Pass@1 & Pass@5 & Pass@10  & Pass@1 & Pass@5 & Pass@10\\
        \midrule
        \multirow{2}{*}{CodeGen} &  Original   &     9.48\% & 19.58\% & 23.48\%      & 13.87\% & 27.12\% & 33.04\%\\
        & Human Label   & 12.26\% & 22.49\% & 25.65\%  & 10.65\% & 21.36\% & 26.52\%\\
        \midrule
        \multirow{2}{*}{\codePLM} &  Original  &   11.83\% & 20.93\% & 27.39\%  & 25.43\% & 37.39\% & 43.04\%\\
        & Human Label  & 13.74\% &21.14\% & 24.78\%  &26.70\% & 40.33\% & 46.09\%\\
        \midrule
        \multirow{2}{*}{ChatGPT} &  Original   & 21.13\% & 27.31\% & 30.00\%   & 35.39\% & 42.77\%& 46.09\%\\
        & Human Label & 26.61\% & 31.31\% & 32.61\%  & 26.96\% & 34.85\% & 37.39\%\\
        \bottomrule
    \end{tabular}
    \vspace{-0.2cm}
\end{table*} 

Fig~\ref{fig.contextual} shows  Acc@K of three models; the results shown in the figure are the average values obtained from 10 experiments. 
We find that the Acc@K value  of the three models is consistent with Pass@K (reflecting that  the generated functions can pass test cases).
To further explore the ability of different models to generate different types of oracle\_context tokens, we divide the oracle\_context tokens into three categories: TypeReference, APIInvocation, and VarReference. Tokens belonging to these three categories typically form the most important part of the target function.
As shown in Fig~\ref{fig.contextual}, on \wenwangbench{} for Python, all models show poor effectiveness in correctly generating the  ``VarReference'' tokens, while the effectiveness in correctly generating the ``TypeReference'' tokens is the best.
However, when it comes to Java, all models have poor effectiveness in correctly generating the  ``TypeReference'' tokens, while the effectiveness in correctly generating ``APIInvocation'' tokens is the best.
The experimental results indicate that the ability of LLMs to generate different types of oracle\_context tokens varies across different languages.
Generating tokens of different categories for different languages is an interesting future research direction.

\begin{framed}
    In summary, as listed in Table~\ref{tab: overall performance for wenwangbench and he new} and Fig~\ref{fig.contextual}, the ability of LLMs to generate the oracle\_context  tokens (reflected by Acc@K) is consistent with that for generating correct code (reflected by Pass@K). As shown in Fig~\ref{fig.contextual}, the ability of LLMs to generate different categories (i.e., TypeReference, APIInvocation, and VariableReference) of oracle\_context tokens varies across different languages.
\end{framed}

\subsection{RQ3: How Do Different Prompts Impact the Effectiveness of These Models?}

We evaluate the three models with both the original docstrings and the human-labeled docstrings, and show the results in Tables~\ref{tab: similarity between human and build-in doc} and~\ref{tab: overall performance for wenwangbench with human label}.
Table~\ref{tab: similarity between human and build-in doc} shows the similarity between the original docstrings and human-labeled docstrings. 
From Table~\ref{tab: similarity between human and build-in doc}, we find that the BLEU-4~\cite{papineni2002bleu} value and Jaccard similarity of the original docstring and human-labeled docstring in a problem of CoderEval for Python is higher than that of CoderEval for Java. 
Table~\ref{tab: overall performance for wenwangbench with human label} shows that for all models, the Pass@K values of the code generated by using the original docstrings and human-labeled docstrings of the three models are more similar on CoderEval for Python than on CoderEval for Java.
On CoderEval for Java, the effectiveness of CodeGen and ChatGPT with the original docstrings  is higher than their effectiveness with the human-labeled docstrings, while the opposite result is observed   when the effectiveness of PanGu-Coder with the original docstrings is compared with its effectiveness with the human-labeled docstrings. 

There are two main reasons for the preceding results. First, the similarity between the original docstrings and human-labeled docstrings for Python is higher than the similarity between the original docstrings and human-labeled docstrings for Java. Second, comparing the generated code with two docstring versions, we find model effectiveness is related to the proportion of the target-language corpus in the whole training data. For Python, high model effectiveness can be attributed to dedicated training or fine-tuning on the Python-code corpus, while for Java, only PanGu-Coder undergoes pre-training on the Java-code corpus. Therefore, while multilingual pre-training can transfer knowledge and bring better generalization across languages~\cite{li2023starcoder}, monolingual pre-training or fine-tuning is still necessary for the model effectiveness.

\begin{framed}
    In summary, we find that the choice of using the human-labeled
docstring vs. the original docstring in a problem has an impact on code
generation effectiveness. In the task of code generation for a single language, when given a docstring with the same semantics but different expressions, the model trained with a single-language corpus performs better than the model trained with a multiple-language corpus. 
 
\end{framed}

\section{related work}\label{sec:related}
HumanEval is a benchmark to evaluate code generation models on the functional correctness of code generated from docstrings~\cite{codex}. It consists of 164 hand-written programming problems (each problem's functionality is reflected by a docstring) along with solutions (in Python), each of which includes a function signature, body, and multiple unit tests.

Following HumanEval, AiXBench~\cite{hao2022aixbench} is proposed to benchmark code generation models for Java. AiXBench contains 175 problems for automated evaluation and 161 problems for manual evaluation. 
The authors of AiXBench present a new metric for automatically assessing the correctness of the generated code, and a set of criteria to manually evaluate the overall quality of the generated code.

MultiPL-E~\cite{Federico2022multibench} is the first multi-language parallel benchmark for text-to-code generation. MultiPL-E extends HumanEval and MBPP~\cite{austin2021mbpp} to support 18 programming languages.
MultiPL-E includes a suite of compilers and an evaluation framework for translating code generation benchmarks (including unit test cases, docstrings, Python-specific terminology, and type annotations) from Python into other programming languages. 
MultiPL-E contains two parallel benchmark portions (i.e., HumanEval and MBPP) for code generation in 18 languages encompassing various programming paradigms, language features, and popularity levels.

While all the preceding benchmarks contain only standalone functions, another benchmark named DS-1000~\cite{lai2022ds1000} contains non-standalone functions. In particular, DS-1000 contains 1000 problems, covering seven widely used Python data science libraries: NumPy, Pandas, TensorFlow, PyTorch, Scipy, Scikit-learn, and Matplotlib.
The authors of DS-1000 mitigate the problem of data leakage by manually modifying functions while emphasizing the use of data from real development scenarios to construct DS-1000. 

Although DS-1000 contains non-standalone functions, DS-1000 has two major limitations. 
First, the functions in DS-1000 are collected from only seven data science third-party libraries in Python.
Second, although some functions in DS-1000 have contextual dependencies on third-party libraries, the development scenarios reflected by DS-1000 are far from pragmatic code generation because 
no function in DS-1000 has  contextual dependencies on user-defined functions outside of the target function. 

Concode~\cite{iyer2018concode} is a new large dataset that contains over 100,000 problems of Java classes from open-source projects. 
From a public Java project on GitHub, the authors of Concode collect each function that has at least one contextual dependency as well as the tuple of natural language (i.e., Javadoc-style method annotations) and code. 
The authors of Concode collect Java functions from approximately 33,000 repositories and then split them into training, validation, and test sets at the granularity of repositories, instead of the granularity of functions.
Although the functions in Concode contain contextual dependencies, Concode uses only BLEU as its  evaluation metric, and does not evaluate the correctness of the generated functions, given that each problem in Concode does not include test cases.

\section{Threats to Validity}\label{sec:threats}

A threat to validity  includes the degree to which the models used in our experiments are representative of true practice. 
All three models used in the experiments come from industry, and each model achieved SOTA performance on HumanEval when it was first proposed.
Due to our limited computing resources at the time of the paper writing, we do not conduct experiments with a number of other recent models such as CodeGeex, StarCoder, and WizardCoder on CoderEval. 
Such threat could
be reduced by more experiments on wider types of subjects
in future work. 
Another threat in our work comes from the accuracy of the statistics for the proportion of standalone vs. non-standalone functions. Section~\ref{context identify} shows the way that we use to count the proportion of standalone vs. non-standalone functions. To identify whether a function has a contextual dependency on a third-party library, we collect all third-party libraries from $pypi$ for Python and $maven$ for Java. Due to our inability to guarantee the collection of all third-party libraries, there may be a slight deviation in the resulting statistics. 
%
%

\section{Conclusion}\label{sec:conclusion}
In this paper, we have presented a new benchmark named \wenwangbench{} to evaluate a model's effectiveness in pragmatic code generation scenarios.
Compared with the widely used HumanEval benchmark, \wenwangbench{} includes carefully selected problems from various open-source projects to evaluate the effectiveness of models in pragmatic code generation. 
The experimental  results show that \wenwangbench{} can reveal the strengths and weaknesses of three LLMs, highlighting the limitations of these existing code generation models in generating non-standalone functions. 
\begin{acks}
This work was partially supported by National Key Research and Development Program Project (2021YFF0704202), National Natural Science Foundation of China under Grant No. 62161146003,  a grant from Huawei, and the Tencent Foundation/XPLORER PRIZE.

We would like to thank Jiaming Huang, Tianyu Chen, Ziyue Hua, and Kun Zhou for their help with improving the quality of unit test cases in CoderEval.
\end{acks}
\balance
\bibliographystyle{ACM-Reference-Format}
\bibliography{sample-base}

\appendix

\end{document}